\documentclass[a4paper,11pt]{article}
\usepackage{jcappub} 
\usepackage{txfonts}
\usepackage{mathtools}
\title{\boldmath Observational tests in scale invariance II: gravitational lensing}

\author[1]{Andr\'e Maeder,}
\author[ 2, 3, 4]{Fr\'ed\'eric Courbin}
\affiliation[1]{Geneva Observatory - chemin des Maillettes 51, CH-1290 Sauverny, Switzerland}
\affiliation[2]{Institute of Physics, Laboratory of Astrophysics, Ecole Polytechnique F\'ed\'erale de Lausanne (EPFL), Observatoire de Sauverny, 1290 Versoix, Switzerland}
\affiliation[3]{ICC-UB Institut de Ci\`encies del Cosmos, University of Barcelona,
Mart\'i Franqu\`es, 1, E-08028 Barcelona, Spain}
\affiliation[4]{ICREA, Pg. Llu\'is Companys 23, Barcelona, E-08010, Spain}

\emailAdd{Andre.Maeder@UniGe.ch}

\abstract{We study the path of light rays passing near a massive object, in the context of the scale invariant equation of the geodesics first obtained by Dirac (1973). Using   the Kottler metric,  we derive the complete equations of the geodesics in the scale invariant context. We find that scale invariance introduces two additional terms to the Einstein term producing the deflection angle and that can potentially act over cosmological distances. Numerical integration of the scale-invariant geodesics, for the specific case of the $z_L \cong 2.0$ lens galaxy in the extreme system JWST-ER1 (van Dokkum et al. 2023; Mercier et al. 2024) shows that the two additional terms  introduce only negligible effects, typically $\mathcal{O}$($10^{-6}$) of the Einstein term. We conclude that the lensing deflection angle derived in Einstein's General Relativity is essentially independent of the scale invariant effects and that the photon's geodesics remain unchanged. 
We also explore the possible origin of the differences in the mass estimates from lensing and photometry 
in JWST-ER1 and in the SLACS galaxies, differences which appear larger at higher redshifts.   Scale invariance appears to release or even suppress the need for dark matter in explaining basic lensing properties.}

\begin{document}
\maketitle
\flushbottom

\section{Introduction} \label{intro}

Gravitational lensing  is a powerful natural phenomenon featuring a vast range of astrophysical and cosmological applications. Its major strength is that it gives access to quantities that are notoriously hard to measure otherwise, masses and mass distributions at all spatial scales being the main ones. Gravitational lensing is also one of the key observations that confirmed General Relativity (GR), through the measurement of the light deflection angle by the Sun during the 1919 Solar eclipse \cite{Eddington19}. It therefore consists in a perfect laboratory to test alternative scenarios to $\Lambda$CDM, such as scale invariant theory \cite{Maeder17a,Maeder23,MaedGueor23}. 
 It is an extension of GR and thus a higher-level theory from the start, by including one more invariance in addition to the general covariance.
As one example application, it can be used  to describe cosmology on the largest scales, in space and time.

Further tests are particularly useful in light of the already performed   observational tests of scale invariant theory, (see paper I \cite{Maeder24a}). It explains: - why the standard mass estimates from galaxy velocities are overestimating the mass of galaxy clusters, - flat rotation curves of galaxies  \cite{Maeder17c}, - why the observed projected velocities of wide binary stars are overestimated by standard theory as their angular separation increases \cite{Hernandez22,MCourbin24c}. Other effects are mentioned in Paper I. Clearly, some effects may be explained equally well by $\Lambda$CDM with dark matter and scale invariance. However, there also are 
effects accounted by scale invariance that are not accounted for by dark matter \cite{Maeder24a}. A typical case is the omnipresent relation between the assumed dark matter (DM) and the baryons first
observed by Lelli et al. \cite{Lelli17} in galaxies of various types and at various distances from their center. The same kind of relation betwen DM and baryons was 
also observed in clusters of galaxies \cite{Chan22} and in highly redshifted galaxies at the peak of their star formation \cite{Nestor Shachar23}. 

In the present work, we confront scale invariant theory to strong gravitation lensing observations at galaxy scale. We compute the geodesic equations in the scale invariance context and show that they do not differ from the standard GR derivation. While scale invariance has no effect on the photons' geodesics themselves, we show that it has an effect on the interpretation and use of the mass-luminosity relation of the stellar population in the lens galaxy and of its IMF.
We illustrate this with real observations of strongly lensed galaxies, focusing on the extreme case of the "dark lens" JWST-ER1 \cite{Dokkum23} at $z_L=1.94$ and on the text-book SLACS sample of lenses at $z_L$between 0 and 0.5 for which excellent spectroscopic observations are available  \cite{Auger09}. This allows us to confront the standard lensing masses at very different redshifts to the luminous masses in the scale invariant framework. 

Section~\ref{geodlight} describes the calculation of trajectories of lensed light paths in the scale invariant context. Section~\ref{SIVlens} shows from accurate numerical integration that scale invariant effects on the light rays are negligible. Section~\ref{lum} examines the origin of the discrepancy between the stellar masses and the masses from the lensing and exposes the scale invariant effects that reconcile the two mass estimates. These are further discussed in  Section~\ref{discussion} which also proposes additional future tests involving gravitational lensing. Finally,  Section~\ref{concl} gives our conclusions.

\section{The lensing effect: the geodesic of deflected light rays}  \label{geodlight}

\citet{Weyl23} and \citet{Eddington23} considered scale invariance  in an attempt to account for
electromagnetism by a geometrical property of space-time.  The original proposition was  abandonned because \citet{Einstein18}
had shown that the properties of a particle would  depend on its past world line, so that 
 atoms  in an electromagnetic field would not show sharp lines. However, since
that time  there has been a number of ways to apply or refer to Weyl invariance.
In recent years, scale invariance has been often investigated in the context of theoretical physics, e.g. 
in Einstein-Cartan gravity with possible implications for fermionic dark matter and properties of the Higgs boson, see ref. \cite{Karananas21}.  
Noticeably, Weyl invariance reduces the number of terms in the action.
We also point out  that there is variety  of scale problems in Astrophysics. For example, regarding the distribution of matter density in the Universe, Einasto et al. \cite{Einasto89} provided evidences of self-similarity of structures on various scales indicating the fractal nature of their distribution.

Here, we refer to the essential works by \citet{Dirac73} and \citet{Canuto77}, who 
considered an interesting new possibility: the Weyl Integrable Geometry (WIG),  where Einstein's criticism does not apply. The condition is that
the coefficient of metrical connection $\kappa_{\nu}$, which expresses the changes of the length $\ell$ of a vector for a small displacement 
 $\delta x^{\nu}$ in this geometry 
should be a perfect differential. For further information, we refer
to the recent work by Maeder and Gueorguiev \cite{MaedGueor23}. There, the general scale invariant field equation and geodesics are demonstrated
from an action principle, which noticeably also determine the gauging condition  of the WIG space. 
The most important conclusion is that some limited scale invariant effects may be present in Universe models with mean density $\overline{\varrho}$
below the   critical value  $\varrho_{\mathrm{c}} = \frac{3 H^2_0}{8 \pi G}$, while scale invariance is killed 
for models with $\overline{\varrho} \geq \varrho_{\mathrm{c}} $.


To calculate the modified trajectories of light rays near a mass, the starting 
point is  the equation of the geodesic \cite{Dirac73,BouvierM78} 
\begin{equation}
\frac{du^{\alpha}}{ds}+ \Gamma^{\alpha}_{\mu \nu} u^{\mu} u^{\nu} -\kappa_{\mu}u^{\mu} u^{\alpha}+ \kappa^{\alpha} = 0.
\label{geod}
\end{equation}
For the metric  of the system, we adopt   
the Kottler metric \cite{Kottler18},
\begin{equation}
ds^2= \frac{-dr^2}{1-\frac{2 GM}{rc^2}-\frac{\Lambda r^2}{3}} -r^2{d\vartheta^2+\sin^2 \vartheta d\varphi^2}+
\left(1-\frac{2 GM}{r c^2}-\frac{\Lambda r^2}{3}\right) c^2 dt^2,
\label{Sch}
\end{equation}
with  $c$ being explicited here.
In the scale invariant system,   $\Lambda = \lambda^2 \Lambda_{\mathrm{E}}$  where $\Lambda_{\mathrm{E}}$ 
is the cosmological constant in the GR  context.  

\subsection{Scale invariant equations of motion in polar ccordinates}

The above  equation of geodesics differs from the classical case of GR by the terms  $ -\kappa_{\mu}u^{\mu} u^{\alpha}+ \kappa^{\alpha}$.
These terms are for the different $\alpha$-coordinates,
%
%
%
\begin{eqnarray}
\alpha =1 &:& \quad -\kappa_{\mu}u^{\mu} \frac{dr}{ds} \quad \rightarrow \quad -\kappa_0 \frac{dt}{ds} \frac{dr}{ds}, \\ \nonumber
\alpha =2 &:& \quad -\kappa_{\mu}u^{\mu} \frac{d\theta}{ds} \quad \rightarrow \quad  -\kappa_0 \frac{dt}{ds} \frac{d\theta}{ds}, \\ \nonumber
 \; \alpha =3 &:& \quad -\kappa_{\mu}u^{\mu} \frac{d\phi}{ds} \quad \rightarrow \quad -\kappa_0 \frac{dt}{ds} \frac{d\phi}{ds}, \\ \nonumber
 \alpha =0 &:&  \quad -\kappa_{\mu}u^{\mu} \frac{dt}{ds} + \kappa_0 \quad \rightarrow \quad  -\kappa_0 \frac{dt}{ds} \frac{dt}{ds}+\frac{\kappa_0}{c^2}. 
\end{eqnarray} 
We recall that $\kappa_0= -\dot{\lambda}/\lambda= 1/t$ and we now  write $c$ explicitely.
The above terms are to be added the classical geodesic equation. We adopt the well known simplified form of the metric,
\begin{equation}
ds^2 \, = \, -e^{\lambda} dr^2-r^2 (d\theta^2+\sin^2 \theta d\phi^2)+e^{\nu} c^2  dt^2,
\label{sc}
\end{equation} 
and have the correspondences,
\begin{equation}
e^{\lambda}=  \frac{1}{1-\frac{2 m}{r}-\frac{\Lambda r^2}{3}}, \quad \mathrm{and} \; \;
 e^{\nu}=\left(1-\frac{2 m}{r}-\frac{\Lambda r^2}{3}\right) ,
\label{defln}
\end{equation}
where  we write $m= GM/(c^2)$, this quantity has the units of a length and it is subject to time-variations \cite{MaedGueor23}.
The above terms  additionally appear to the classical expressions of the components of the geodesics,
\citep{Tolman34,Weinberg72}. 
The general equation we are searching could be applied to a variety of post-Newtonian problems in addition to lensing, such as the advance of the perihelion, etc.
Thus, we will impose the specific conditions  for the propagation of light rays only at the very end, on the general equation. This is also the reason 
 why the final equations will be expressed for the inverse radius $u=1/r$  depending on angle $\phi$, 
in order to correspond to the  usual Binet equation, which in classical Mechanics provides the form of the planetary orbits
for a given force law, and vice-versa.  
We are also supposing  
that the motions occur in a plane with a constant value of $\theta=\pi/2$, and get
%
%
%
\begin{eqnarray}
\frac{d^2 r}{ds^2}+\frac{1}{2} \frac{d\lambda}{dr}\left(\frac{dr}{ds}\right)^2- r e^{-\lambda}\left(\frac{d\phi}{ds}\right)^2 +
\frac{e^{\nu-\lambda}}{2}\frac{d\nu}{dr}\left(\frac{c^2 \, dt}{ds}\right)^2  
-\kappa_0 \frac{dt}{ds} \frac{dr}{ds} &=& 0,  \label{d2r} \\
\frac{d^2\phi}{ds^2}+\frac{2}{r} \frac{dr}{ds} \frac{d\phi}{ds}  -\kappa_0 \frac{dt}{ds} \frac{d\phi}{ds} & = & 0,  \label{eph}\\
\frac{d^2 t}{ds^2}+ \frac{d \nu}{ds}\frac{dt}{ds}-\kappa_0 \left(\frac{dt}{ds}\right)^2+\frac{\kappa_0}{c^2} & = & 0, \label{et}
\end{eqnarray}
%
where we may verify the consistency of the dimensions.
The last two equations may be easily integrated. Let us call $\upsilon= \frac{d\phi}{ds}$, getting for (\ref{eph}),
\begin{equation}
\frac{d\upsilon}{ds}+ \frac{2}{r} \frac{dr}{ds} \upsilon -\frac{1}{t}\frac{dt}{ds} \upsilon=0, \quad \mathrm{and} \; \;
\frac{d\upsilon}{\upsilon}+ 2 \frac{dr}{r}- \frac{dt}{t}=0.
\end{equation}
This gives 
\begin{equation}
r^2 \, \frac{d\phi}{ds} = h t,
\label{momang}
\end{equation}
where $h$ is an integration constant,  its meaning is the angular momentum per mass. This last expression is the 
modified conservation law of the angular momentum. 
We see that it increases with time $t$, a behaviour which is quite consistent with that in the scale invariant two-body problem
\citep{MBouvier79,Maeder17c}, where there is a slow secular increase of the orbital radius of a body. At the same 
time, the orbiting body  is keeping a constant orbital velocity $r \, \frac{d\phi}{ds}$  during its evolution. 
These well established secular effects of scale invariance
are always very small  in our current Universe, since they are globally scaling with the inverse of the age of the Universe.
Turning to the third equation,  we first multiply it by $\left(\frac{ds}{dt}\right)^2$ and get,
\begin{equation}
\frac{d^2t}{dt^2}+\frac{d\nu}{dt} \, \frac{dt}{dt}- \kappa_0+\frac{\kappa_0}{c^2} \, \left(\frac{ds}{dt}\right)^2= 0.
\end{equation}
The first two terms are zero, the second one because $\nu$ is a scale invariant quantity, independent of time,  which is easily verified  
by considering the terms in the parenthesis of expression (\ref{Sch}): 
 both the mass and radius are scaling
like $1/\lambda$,  so that $2m/r$ is scale invariant;  $\Lambda$ is scaling like $\lambda^2$ so that $(\Lambda r^2)/3$ is also
scale invariant.  
This means that the block of the two terms with $\kappa_0$ coming from scale invariance is also equal to zero  in the third equation. 
Thus,  regarding  the time component of the field equation, one is just brought back to the classical case and 
 calling  $w=ds/dt$,  one is left to,
\begin{equation}
\frac{dw}{ds}+  \frac{d\nu}{ds} w =0,
\end{equation}
with a solution $w=dt/ds= k \, e^{-\nu(s)}$, where $k$ is a constant \footnote{We note that instead of this solution, we could take the condition 
 $ds/dt=c$ set by the block of terms containing $\kappa_0$.  When we  do it in the case of 
the geodesic of light rays, we find the same final equation (\ref{BM}), which is consistent since the condition $ds=c dt$ is trivially verified by photons.
This shows the coherence of the overall description.}.

Now,  we turn to the more difficult Eq.~(\ref{d2r}). In this respect, let us quote \citet{Eddington23} who said
about the corresponding  equation in the GR context:{\emph{
"Instead of troubling to integrate (\ref{d2r}) we can use in place of it  (\ref{sc})}}", (here the labels of both equations 
have been changed appropriately). 
Thus from (\ref{sc}) we get,
\begin{equation}
e^{\lambda} \left(\frac{dr}{ds}\right)^2 + r^2  \left(\frac{d\phi}{ds}\right)^2 - 
e^{\nu} c^2 \left(\frac{dt}{ds}\right)^2 +1=0.
\end{equation}
With the above expression for $w$, the third term can also be written 
\begin{equation}
-e^{\nu} c^2 k^2 e^{-2\nu}=  -c^2 k^2 e^{-\nu}=
 -\frac{c^2 k^2}{[ 1-\frac{2 GM}{r c^2}-\frac{\Lambda r^2}{3}]}.
\end{equation}
Thus, the above equation becomes,
\begin{eqnarray}
\left(\frac{dr}{ds}\right)^2 +  r^2  \left(\frac{d\phi}{ds}\right)^2  -\frac{2m}{r} \left(1+ r^2  (\frac{d\phi}{ds})^2 \right) 
-\frac{\Lambda r^2}{3}\left(1+ r^2  (\frac{d\phi}{ds})^2\right)= c^2 k^2-1.
\label{B1}
\end{eqnarray}
Eliminating $(\frac{d\phi}{ds})$ with (\ref{momang}), we get
\begin{equation}
\left(\frac{ht}{r^2} \frac{dr}{d\phi}\right)^2 +\frac{h^2 t^2}{r^2}=c^2k^2 -1+ \frac{2m}{r}+ 
\frac{2mh^2 t^2}{r^3}+\frac{\Lambda r^2}{3} +\frac{\Lambda    h^2 t^2}{3}.
\end{equation}
We divide this equation by $h^2 t^2$ and then have an equation analogous to the Binet 
equation which describes the planetary orbit in polar coordinates by setting
$u= 1/r$,
\begin{equation}
\left(\frac{du}{d\phi} \right)^2 +u^2 =\frac{c^2 k^2 -1}{h^2 t^2}+ \frac{2mu}{h^2 t^2}+ 
2 mu^3 +\frac{\Lambda}{3 u^2 h^2 t^2} +\frac{\Lambda}{3}.
\end{equation}
Now,  we derive this equation with respect to $\phi$ and obtain,
%
%
%
\begin{eqnarray}
2 \left(\frac{du}{d\phi}\right) \frac{d^2u}{d\phi^2}  +2 u \frac{du}{d\phi} & = &
\frac{c^2 k^2 -1}{ h^2} \frac{d}{d\phi} \left(\frac{1}{t^2}\right)  + \frac{2m}{h^2 t^2}  \frac{du}{d\phi} \nonumber\\
& + &\frac{2u}{h^2 t^2} \frac{dm}{d\phi}+   \frac{2mu}{h^2} \frac{d}{d\phi}\left(\frac{1}{t^2} \right) +
6 mu^2 \frac{du}{d\phi}+2 u^3 \frac{dm}{d \phi} \nonumber   \\
&-& 2 \frac{\Lambda_{\mathrm{E}} \lambda^2}{3  h^2 t^2} \frac{1}{u^3}\frac{du}{d\phi}+
\frac{\Lambda_{\mathrm{E}} }{3  h^2 u^2} \frac{d}{d\phi} \left(\frac{\lambda^2}{t^2}\right)
+\frac{2\Lambda_{\mathrm{E}} \lambda}{3} \frac{d\lambda}{d\phi}.
\end{eqnarray}
There, we have  expressed $\Lambda= \Lambda_{\mathrm{E}} \lambda^2$ \cite{Canuto77}, \cite{Maeder17a}.
There is also a direct relation between $\phi$ and time since  
the time $t$ is uniquely defined  along the geodesic described by the variation of angle $\phi$, see Fig. \ref{Basic}. 
Thus, $\lambda$  which is a function of time $t$
is also a  function of $\phi$.
Let us now multiply the result by $\frac{1}{2} \frac{d \phi}{du}$,
%
%
%
\begin{eqnarray}
\frac{d^2u}{d\phi^2}  +u & = &\frac{c^2 k^2 -1}{2 h^2} \frac{d}{du} \left(\frac{1}{t^2}\right)  + \frac{m}{h^2 t^2}+ \frac{u}{h^2 t^2} \frac{dm}{du}+
\frac{mu}{h^2} \frac{d}{du}\left(\frac{1}{t^2} \right) \nonumber \\
& + &3 mu^2+u^3 \frac{dm}{du} -\frac{\Lambda_{\mathrm{E}} \lambda^2}{3  h^2 t^2} \frac{1}{u^3}+
\frac{\Lambda_{\mathrm{E}} }{6  h^2 u^2} \frac{d}{du} \left(\frac{\lambda^2}{t^2}\right)
+\frac{\Lambda_{\mathrm{E}} \lambda}{3} \frac{d\lambda}{du},
\label{compl}
\end{eqnarray}
The terms containing $ \Lambda_{\mathrm{E}}$ and the derivatives of $m$ are currently not present.
This equation is the basic one combining the effects of General Relativity and scale invariance 
to study the advance of the perihelion of planets, the light deflection by a massive body as well
as the gravitational shift of spectral lines. In the Newtonian approximation, it
gives the current Binet equation of planetary orbits in polar coordinates \citep{BouvierM78}.

\subsection{The geodesics  of  light rays}

The trajectory of a light ray alike that of a free particle is governed by the equation of geodesics, however
with the additional condition  $ds=0$, {\emph{i.e.}} its proper time does never change.
 According to Eq.~(\ref{momang}), this implies $h=\infty$, also meaning that the angular 
momentum associated to a light ray cannot be considered as a finite quantity. This
is leading to major simplifications in the above equation (\ref{compl}), 
where only 3 terms are remaining on the right side.
From the gauging condition   $3 \,\frac{ \dot{\lambda}^2}{\lambda^2} \, =
\,c^2 \lambda^2 \,\Lambda_{\mathrm{E}} $, adopted in \cite{Maeder17a} and further supported by an action principle  \cite{MaedGueor23} (with a 
term $c^2$ for the consistency of the units),  we got with $\lambda=t_0/t$ in the units of the cosmological models  with
$\Lambda_{\mathrm{E}} = \frac{3}{c^2 \, t^2_0}$,  and in current units
$\Lambda_{\mathrm{E}}=  \frac{3}{c^2 \, \tau^2_0}$ with $\tau_0$ the age of the Universe, see \cite{MaedGueor21a}. Thus, Eq.(\ref{compl}) becomes
\begin{equation}
 \frac{d^2u}{d\phi^2}  +u =  3 mu^2+u^3 \frac{dm}{du}+ \left(\frac{1}{c^2   \tau^2_0}\right) \lambda \frac{d\lambda}{du}.
\label{BM}
\end{equation}
 The scale factor $\lambda$ is subject to some limitations, since $\lambda=t_0/t$, and $t$ varies between 
 $t_{\mathrm{in}}= \Omega^{1/3}_{\mathrm{m}}$ and $t_0=1$.  
 One is noting that if
$\lambda$ is a constant, we again have the  fundamental   equation for lensing in GR, since $dm$ also disappears.
Quantity   $c \tau_0$ is a distance  in Gyr (or seconds) equal to the age the Universe, 
it is  of the order of the radius of the Hubble sphere $c/H_0$.

We recall that the observable Universe has a different definition
 given by the particle horizon $d_{\mathrm{H}}$ \citep{Rindler56, Rindler69}. 
It has been verified that SIV models with $\Omega_{\mathrm{m}} < 1$
do have a particle horizon \citep{MaedGueor21a}. The  different
estimates of the size of the observable Universe are of  the same order of magnitude as $c \tau_0$,
although their formal definitions are different.
 For example in the EdS model $d_{\mathrm{H}}=3 c \tau_0$. There, 
the very initial expansion, before a further global braking, was much faster than the present one, thus a numerical factor 
larger than 1.
The SIV models show an initial phase of moderate braking
followed by a progressive acceleration phase \citep{Maeder17a}, thus their $d_{\mathrm{H}}$ are close to $c \tau_0$.
Thus, we may  consider   $c \tau_0$ as the  order of magnitude of the present "size" of the connected Universe. 

\subsubsection{The classical cases}

In absence of the second member of equation (\ref{BM}), 
\begin{equation}
\frac{d^2u}{d\phi^2}  +u =0, \quad  \mathrm{one \; has}   \; \; u_0 \, = \, \frac{\cos  \phi}{R}
\label{droite}
\end{equation}
as a solution,
 where $R$ is the impact parameter (Fig.~\ref{Basic}). With this solution, the two terms on the left side cancel each 
other.
When $\phi$ is  increasing from zero $ \rightarrow \pi/2$, a straight line is described (cf. Fig.~\ref{Basic}),
 as expected for a light ray in absence of a massive body.

The terms on the right of (\ref{BM}) are generally very small, in particular the last two, as we shall below.
The usual treatment \citep{Eddington23} for the first term is to introduce the solution (\ref{droite}) 
into $3 m u^2 $ and to solve the equation,
\begin{equation}
\frac{d^2u}{d\phi^2}  +u =  \frac{3 m}{R^2} \cos^2 \phi.
\label{BM1}
\end{equation}
 A particular solution of this equation is $u_1= \frac{m}{R^2} \left(\cos^2 \phi+2 \sin^2 \phi \right)$,
so that a good  approximation for the solution of (\ref{BM1}) is,
\begin{equation}
u  \, = u_0+u_1=     \, \frac{\cos \phi}{R}+ \frac{m}{R^2} \left(\cos^2 \phi+2 \sin^2 \phi \right).
\label{droite1}
\end{equation}
This is the usual basic expression for the path of the deflected light rays  derived from GR.
The deflection by the GR effect is essentially  produced within
a limited distance from the lens center (a few impact parameters $R$).
The well known solution for the deflection angle $\hat{\alpha}_{\mathrm{GR}}$   is,
\begin{equation}
\hat{\alpha}_{\mathrm{GR}} = 4 \frac{m}{R} = \frac{4\, GM}{c^2 R}   \quad \mathrm{[rad]} , 
\label{dgr}
\end{equation} 
where $M$ is the mass of the deflector and $R$ the impact parameter.


\section{The scale invariant effects in lensing} \label{SIVlens}

The term $GM/R$ in the above equation is the gravitational potential which as we have seen is an invariant quantity. Thus we might wonder whether in the scale invariant context the deflection angle  should not just be unchanged. Indeed, despite the fact that the potential remains the same, Eqs.~(\ref{compl}) and (\ref{BM}) contain additional terms to the Einsteinian one. Also, the equation of motion in the Newtonian approximation is modified with an additional term, potentially significant in the astronomical context. Thus it makes sense to wonder whether the path of a light ray also undergoes some changes with respect to GR.

A difficulty of  Eq.~(\ref{BM}) is that, in addition to the independent variable $\phi$, it requires the variable $t$  to express the term $\lambda$. The two variables are related  by geometrical relations on the light path. However, the result is an intricate equation, making numerical integration mandatory to evaluate relative importance of the additional terms and their impact on the deflection of light rays.

\subsection{The SIV relation between redshifts and ages}   \label{agez}
 
To proceed to numerical integration of Eq.~(\ref{BM}), we need some basic relations  between  the observed redshift $z$  
and time $t$, which then gives the scale invariance factor $\lambda=1/t$. The cosmological equations were derived from the
scale invariant field equation with the current FLWR metric \cite{Maeder17a}; an analytical solution for the flat model has been obtained by Jesus \cite{Jesus18},
\begin{equation}
a(t)= \left[\frac{t^3-\Omega_{\mathrm{m}}}{1-\Omega_{\mathrm{m}}}\right]^{2/3},  \quad \mathrm{with}
 \;\; H(t)= \frac{2\, t^2}{t^3-\Omega_{\mathrm{m}}}.
\label{at}
\end{equation} 
There, the units are such that  at present time $t_0 =1$ one has $a(t_0)=1$, while at the origin $a(t_{\mathrm{in}})=0$ one has 
$t_{\mathrm{in}}= \Omega^{1/3}_{\mathrm{m}}$.
With the FLWR metric, the current relation $ 1+z = \frac{a(t_0)}{a(t)}$ between the redshift $z$
 and the expansion factor $a(t)$ is preserved. 
This  is  leading to,
\begin{equation}
t= \left[\Omega_{\mathrm{m}}+(1+z)^{-3/2}(1-\Omega_{\mathrm{m}})\right]^{1/3}.
\label{tz}
\end{equation} 
The linear  relation between the  ages in the above $t$ scale and $\tau$ in the current units    is evidently,
$\frac{\tau - \tau_{\mathrm{in}}}{\tau_0 - \tau_{\mathrm{in}}} = \frac{t - t_{\mathrm{in}}}{t_0 - t_{\mathrm{in}}}$.
This gives with $\tau_0$ the present age (13.8 Gyr),
\begin{eqnarray}
\tau = \tau_0 \, \frac{t- t_{\mathrm{in}}}{1- t_{\mathrm{in}}} \,  \;  \mathrm{and} \; \;
t = t_{\mathrm{in}} + \frac{\tau}{\tau_0} (1- t_{\mathrm{in}}) , \;
\mathrm{with} \; \frac{d\tau}{dt} \, = \, \frac{\tau_0}{1-t_{\mathrm{in}}}, 
\label{dt}
\end{eqnarray}
For increasing  $\Omega_{\mathrm{m}}$,  
the timescale $t$  (and thus the $\lambda$-variation!) is squeezed since $t_{\mathrm{in}}=\Omega_{\mathrm{m}}^{1/3}$ is increasing towards 1. 
With $t=\Omega^{1/3}_{\mathrm{m}}+ \frac{\tau}{\tau_0}(1-\Omega^{1/3}_{\mathrm{m}})$, one gets  the age $\tau$ 
in current units for a galaxy of redshift $z$, 
\begin{equation}
\tau = \tau_0 \frac{\left[[\Omega_{\mathrm{m}}+(1+z)^{-3/2}(1-\Omega_{\mathrm{m}})\right]^{1/3}- \Omega^{1/3}_{\mathrm{m}}}
{1- \Omega^{1/3}_{\mathrm{m}}}.
\label{tz2}
\end{equation}
This relation can also be expressed as a function of $H_0$, by combining  with  
$H_0= \frac{2 (1- \Omega^{1/3}_{\mathrm{m}})}{\tau_0 (1-\Omega_{\mathrm{m}})}$ from Eqs. (\ref{at}) and (\ref{dt}), 
\begin{eqnarray}
\tau = \frac{2}{H_0}\frac{\left[[\Omega_{\mathrm{m}}+(1+z)^{-3/2}(1-\Omega_{\mathrm{m}})\right]^{1/3}- \Omega^{1/3}_{\mathrm{m}}}
{1-\Omega_{\mathrm{m}}}. \label{th}
\end{eqnarray}
With these expressions in hand we can now associate an age to any object at any redshift.


\subsection{The variations of the term $dm$} \label{dmm}

During  the path of the light ray  from  the source, $S$, to the observer, $O$, the mass of the  lens undergoes some limited variations due to  the scale variations during the travel time. This results in the   term $u^3 (dm/du)$ in Eq.~(\ref{BM}), coming in addition to the classical  GR term  $3 m u^2$. This is because, on the light path, both $m$ and $u$ increase at the same time, so that $(dm/du)$ is positive. Alike a tidal effect, it behaves  in  $u^3 = 1/r^3$, thus being  essentially located close to the lens, even more than the gravity effect itself. Since $m$ has the units of length (see Eq.~\ref{defln}),  $u^3 (dm/du)$ has the units of the inverse of a length.

Do these extra terms impact the deflection angle in the SIV context? To evaluate the effect, we consider the practical case of the remarkable strongly lensed galaxy JWST-ER1 recently observed by van Dokkum et al.  \cite{Dokkum23} with the JWST. It forms a complete Einstein ring with a radius $\theta_E =1.54 \pm 0.02$ arcseconds produced by a massive quiescent lensing galaxy at $z_L=1.94$  acting on a source at $z_S=2.98$. The total lensing mass is estimated to be $M_{\rm Tot} = 6.5 \times 10^{11}$ M$_{\odot}$ within a radius of 6.6 kpc at the lens redshift, while the stellar mass derived from multi-band photometry is 5.9 times lower, with $M_{\star}=1.1 \times 10^{11}$ M$_{\odot}$.
 Inbetween, an in-depth analysis of JWST-ER1 by Mercier et al. \cite{Mercier24}, based on 25 bands in the visible and near infrared,
has led  to a better  established value of the  redshift of the source galaxy
 $z_{\mathrm{S}}= 5.48\pm0.06$, rather  different from the previous value.
As shown below the exact value of the redshift  of the source has essentially no effect on the light deflection. 
The redshift of the lens is about the same, $z_{\mathrm{L}}=2.02\pm 0.02$, a total mass within the Einstein radius of
$(3.66\pm 0.36) \times 10^{11}$ M$_{\odot}$  and a total 
 stellar mass of $1.37^{+0.14}_{-0.11} \times 10^{11}$ M$_{\odot}$  have been found by these last authors.   
 As stated by van Dokkum et al., this is a beautiful example of a system where {\emph{"Additional mass appears to be needed to explain the lensing results, either in the form of a higher-than-expected dark matter density or a bottom-heavy initial mass function."}}

\begin{figure*}
\centering
\includegraphics[width=15cm, height=7cm]{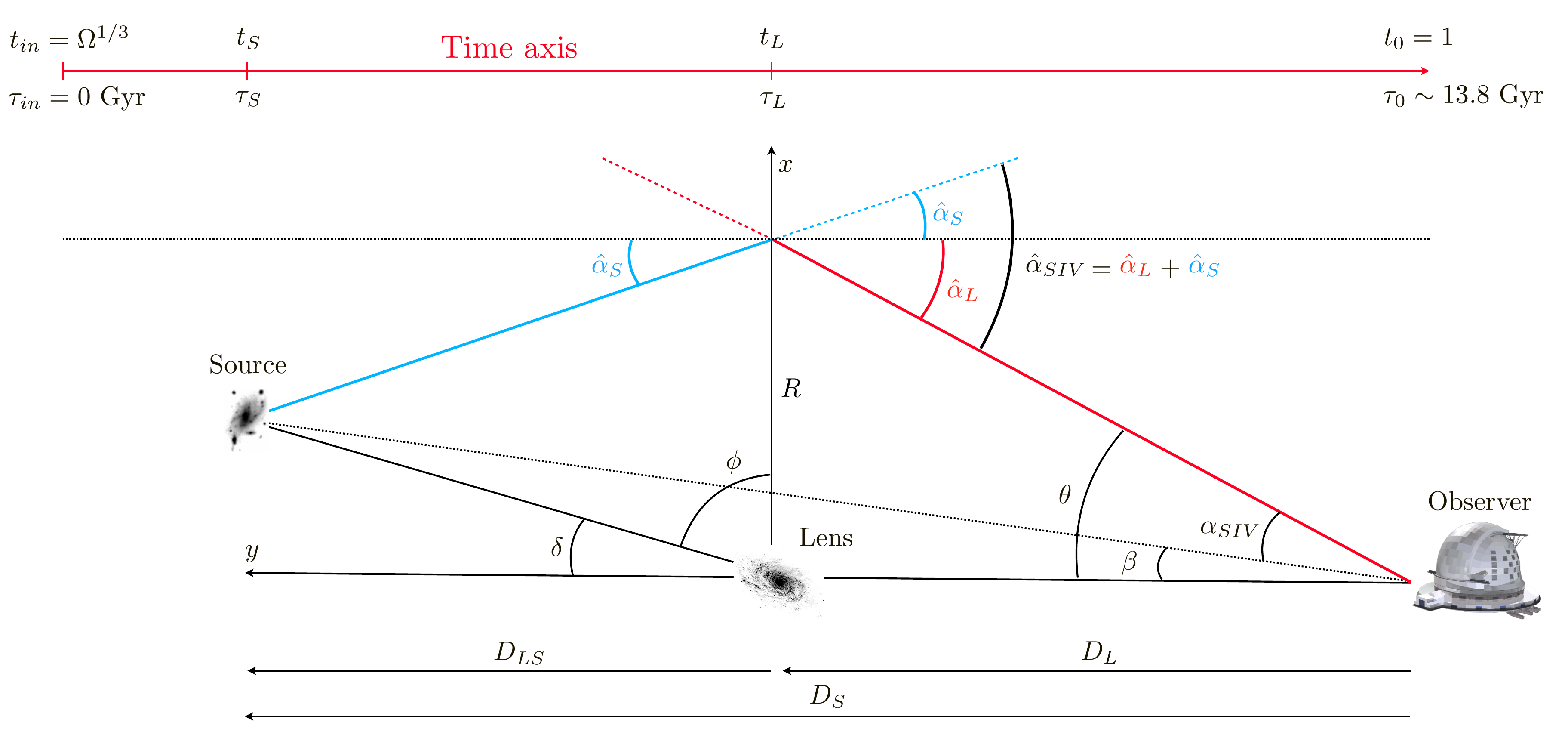}
\caption{The geodesic of a deflected light ray. Basic sketch of the lensing geometry, with the impact parameter, $R$, in the lens plane, the polar coordinates $r$ and $\phi$, the distance $ D_{\mathrm{S}}$ from  the observer to the source, $ D_{\mathrm{L}}$ from  the observer to the lens, $ D_{\mathrm{LS}}$ from the lens to the source and the $\hat{\alpha}$  deflection angles.
 }
\label{Basic}
\end{figure*}

We use JWST-ER1 as a test bench for our integration of the light path. This path is expressed in terms of the variable $u=1/r$ as a function the angle $\phi$,  starting close to $\pi/2$ for the direction pointing to the distant source, S, all the way to the nearest point at the impact parameter where $u= 1/R$ and where $\phi=0$ as can be seen from Fig.~\ref{Basic}. We adopt a step size of $2 \times 10^{-4}$ for the variable $\phi$. Tests made with smaller steps give identical results. The calculations represent an isochrone described by about 7850 points, which ensures a high accuracy of the geodesics. 
From the relation between redshifts and cosmic ages (Sect.~\ref{agez}), we have in the $t$-scale an age of  the source 
given by $t_{\mathrm{S}}= 0.5536$  ($t_{\mathrm{S}}= 0.4756$ with \cite{Mercier24}) 
 and for the lens $t_{\mathrm{L}}= 0.6201$.
We recall that $t$ is dimensionless, with $t_0=1$. (The lens mass $M_{\rm Tot}$ has been taken, as well as a value $\Omega_{\mathrm{m}}=0.05$,
  thus assuming no dark matter.) 
 At each point defined by coordinates $(u, \phi)$  along the path of the light ray, we can  then assign an age $t$ and a value $\lambda$ ($\tau$ is not needed here). We may then obtain  the time variations of the mass as well as the term $dm/du$. In terms of numerical results, we obtain for example that in the specific case of JWST-ER1, the ratio 
\begin{equation}
\mathcal{R} =\frac{u^3 \frac{dm}{du}}{3 mu^2},
\label{rr1}
\end{equation} 
\vskip 1pt
\noindent
is equal to $6.62 \times10^{-6}, \; 2.25 \times 10^{-7}, \; 6.36 \times 10^{-6}$ for angles of $\phi$ = 89, 45 and 1 degree respectively. In other words, the additional  term  $u^3 (dm/du)$ has a totally negligible effect on the resulting deflection angle $\hat{\alpha}_{\mathrm{S}}$.   The fact that the source 
is at $z_{\mathrm{S}} =5.48$  is increasing the path of the light rays, but the  ratio $\mathcal{R}$  always remains very limited. 
This is evident if  we consider the behavior  of ratio,
\begin{equation}
\mathcal{R} \sim \,    u \frac{\Delta m}{\Delta u} \, m \sim \;  \rightarrow 0,
\label{dist}
\end{equation}
which is tending to zero, like $u$ at very large distances. 
Indeed, we have to consider two effects in the changes of distance and mass,
 the changes due to scale invariance and the change due to a larger distance of the source.
For the first effect, the ratio is constant,  since the time dependence of the relative changes  $\frac{\Delta r}{r}$ and 
$\frac{\Delta m}{m}$  are the same  \cite{MaedGueor23}. 
For the second, we note that at large distances  the ratio $\frac{\Delta m}{\Delta u}$ remains  limited.
The Einsteinian term $3mu^2$ is itself strongly decreasing with the distance to the lens, 
 so that the scale invariant term  $u^3 \frac{dm}{du}$  is   very small at large distances. 

 The above tests were made for 
$\Omega_{\mathrm{m}}=0.05$, for any larger value of $\Omega_{\mathrm{m}}$ the variations of $\lambda$ would always be even  smaller and 
the same for  $\frac{dm}{du}$.  Thus, the above conclusion applies for any value of $\Omega_{\mathrm{m}}$.

\subsection{The  $\Lambda$  term  and general remarks} \label{Lam}

We now turn to the other additional term $ \frac{\Lambda_{\mathrm{E}}}{3} \,\lambda \frac{d\lambda}{du}$ in Eq.~(\ref{BM}), 
with $\Lambda_{\mathrm{E}}=  \frac{3}{c^2 \, \tau^2_0}$ in the scale invariant theory, see also \cite{MaedGueor21a}.
 Physically,
the above  term results from the cosmological constant, present in  the Kottler metric  (\ref{Sch}).
 The occurence of this  term is not at all related to the deflecting mass and it is  present anyway, with or  without lensing.
It is  associated to the specific geometry  of the system. In absence of a mass, the metric of the system would  just be the de Sitter metric, 
which is  endowed with a  curvature term. In this context, the above term being always present in the geometry is not
 an additional correction to the classical Einsteinian term. Moreover, it  is likely to be
a very small quantity, as estimated below,  proportional to the cosmological constant. 

The above calculations for JWST-ER1  provide  the information on the magnitude of this effect. 
Although this geometrical term has no effect on the light deflection of any particular mass,  we may compare its value 
 to the classical Einsteinian term $3 m u^2$ , in order to see its relative importance. For example, the  ratio 
$ \left(\frac{\Lambda_{\mathrm{E}}}{3} \,\lambda \frac{d\lambda}{du}\right)/ (3 m u^2)$
is $1.38 \times 10^{-5}$, 
$1.16 \times 10^{-8}$, and $2.30 \times 10^{-7}$ for angles of $\phi = 89, 45$  and 1 degree respectively. 
The  so-called $\Lambda$ term  is thus very  small with respect to lensing effects, on which it has anyway no effect.
The other possible cosmological  implications of this term, such as on the CMB or first evolutionary stages are beyond the scope of this work.


We have studied the nice case of JWST-ER1, but what about lensing by larger masses, such as galaxy clusters? As far as the mass variations are concerned, i.e., $u^3 \frac{dm}{du}$, we see that its ratio to the Einsteinian term $3 mu^2$ changes as $\frac{1}{3} \frac{dm}{m}\frac{u}{du}$.  Thus, as the relative change of the mass $dm/m$ is independent of the mass (since $dm  \sim m$), our conclusion applies to any mass, thus including clusters of galaxies.

The deflection angle in lensing systems  is  essentially independent of scale invariant effects, as shown above.  
      The present detailed analysis demonstrates that the local  inhomogeneity and anisotropy of the space-time created by a mass element in the scale invariant context  introduce  only  minor differences in the light deflection as compared to standard GR.  These  effects are clearly  not sufficient to account for the large  mass difference observed  in JWST-ER1 \citep{Dokkum23} between the masses from the lensing effect
and from the observed luminosity.

This confirms that the masses obtained by the lensing effect are most remarkably 
 independent of scale invariant effects. This result in itself is a fundamental one:
 the geodesics of light in scale-invariance theory do not change with respect to standard GR assumptions. 


\section{The origin of the discrepancy between the stellar and lensing masses}  
\label{lum}

Since the gravitational lensing deflection angle appear to be free from scale invariant effects, we may now turn to the impact of scale invariance on lens mass estimates in the Einstein radius, as derived from photometry or spectroscopy of the lens. 

\subsection{The emblematic case of JWST-ER1s }
 
\begin{figure*}
\centering
\includegraphics[width=15cm, height=8cm]{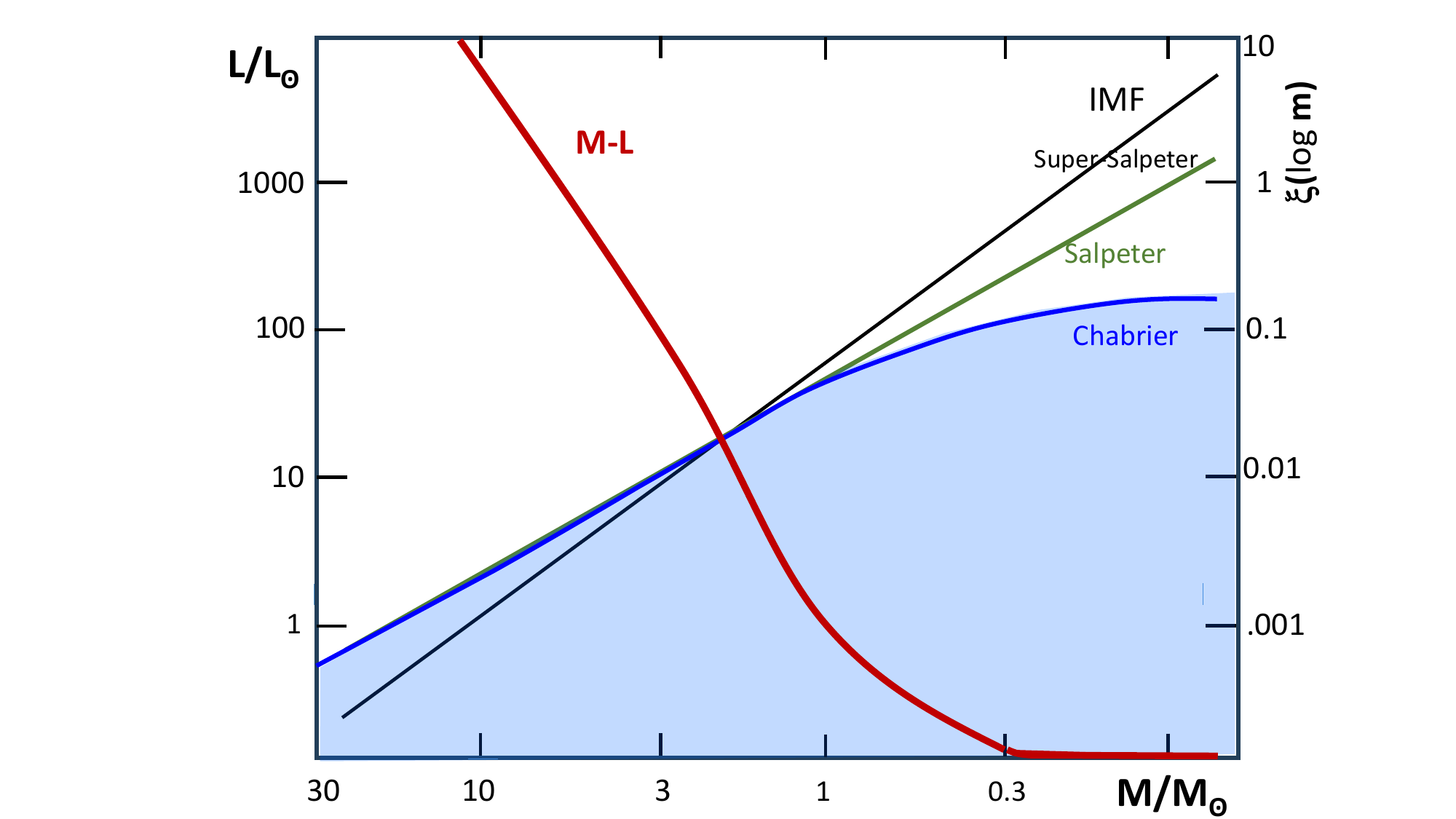}
\caption{The red thick curve represents the mass-luminosity relation on the zero-age sequence \cite{Maeder09}, according to the scale indicated on the left
vertical axis. The other curves describe various IMF:  Chabrier \cite{Chabrier03},  Salpeter with $x=1.35$ \cite{Salpeter55} and    the so-called Super-Salpeter by \cite{Dokkum23},  according to the scale indicated on the right vertical axis. 
Note that in $\xi(\log m)$ the log is decimal. The Figure is inspired from \cite{Dokkum23}.
 }
\label{imf}
\end{figure*}

We consider here again the remarkable case of JWST-ER1 and its high lensing-to-stellar mass ratio of $\sim 6$ according to van Dokkum et al. \cite{Dokkum23} and already presented in Sect.~\ref{dmm}. With a lens redshift  $z_{\mathrm{L}}=1.94$, it is a textbook example of a distant  massive quiescent galaxy with a low star formation rate. It has  an age of $\tau_\mathrm{L} = 1.9 ^{+0.3}_{-0.6}$ Gyr, implying that relatively low mass stars dominate the stellar population of the lensing galaxy. 
Its unitless age is $t_{\mathrm{L}}=0.62$ (cf. Eq.~\ref{tz}), taking  $\Omega_{\mathrm{m}} =0.05$, i.e. assuming  no dark matter
(the Big-Bang occured at $t_{\mathrm{in}}= \Omega^{1/3}_{\mathrm{m}}$, i.e.  0.3684). According to the scaling of masses $M \sim t$ 
\cite{MaedGueor23}, this would  imply that the inertial and gravitational masses in JWST-ER1 is equal to a fraction 0.62 of its present value. 
Incidentally this might also contribute to the observed compactness of the early galaxies.
This means that the stellar mass distribution in JWST-ER1 may be shifted to lower masses with respect to 
today's distribution. Interestingly enough, the consequences of this shift have rather similar effects to those of
 the solution proposed by van Dokkum et al. \cite{Dokkum23}  to solve the mass discrepancy between the mass infered from the lens luminosity
(stellar mass $M_{\star}=1.1 \times 10^{11}$ M$_{\odot}$) and its lensing mass, 
$M_{\mathrm{Tot}} = 6.5^{+3.7}_{-1.5} \times 10^{11} \; M_{\odot}$.
They  suggest a shift from Chabrier's IMF \citep{Chabrier03}, which has relatively  few low-mass stars giving  a mass of $M_* = 1.1^{+0.2}_{-0.3} \times 10^{11} \; M_{\odot}$, to a very bottom-heavy IMF called "Super-Salpeter IMF" giving a mass of $M_* = 4.0^{+0.6}_{-0.8} \times 10^{11} \; M_{\odot}$.  The standard Salpeter IMF gives  an intermediate value of $M_* = 2.0^{+0.5}_{-0.5} \times 10^{11} \; M_{\odot}$. Fig.~\ref{imf} illustrates the different IMF assumptions.

 We mentioned in Sect. \ref{dmm} the new results on JWST-ER1 by Mercier et al. \cite{Mercier24} with in particular a much higher redshift
$z_{\mathrm{S}}= 5.48\pm0.06$ of the source, while the redshift of the lens is about the same  $z_{\mathrm{L}}=2.02\pm 0.02$. The estimated
total and stellar masses from the lensing and photometry are also different, respectively  
$M_{\mathrm{Tot}} = (3.66\pm 0.36) \times 10^{11}$ M$_{\odot}$  and a 
 stellar mass of  $ M_*= 1.37^{+0.14}_{-0.11} \times 10^{11}$ M$_{\odot}$. 
For the source, the shift in time $t$ is  from $t_{\mathrm{S}}= 0.554$  to  0.476 
 and for the lens from  $t_{\mathrm{L}}= 0.620$ to 0.614. The large change of the source redshift has only a negligible effect on the deflection
angle, since as well known it only the region close to the minimum distance which plays a significant role because the Einsteinian effect behaves like
$3 m u^2$ and at large distances $u \rightarrow 0$. The scale invariant effect $u^3 \frac{dm}{du}$ is evidently also totally negligible.

The rather ad-hoc shift produced by the assumption of a Super-Salpeter IMF  leads to a large increase of the mass-to-light ratio since it assumes a relatively larger fraction of lower stellar masses,
with lower luminosities (Fig.~\ref{imf}). 
Even with the Super-Salpeter assumption, a significant fraction of dark matter is still necessary, 
  for both goups,  to explain the observed Einstein radius of this system.

The observed luminosity $L_{\mathrm{obs}}$  reflects the true mass of the stars as they are at any given age. For an age of $\tau_\mathrm{L} = 1.9^{+0.3}_{-0.6}$ Gyr assigned by Prospector{\footnote{ PROSPECTOR, a flexible code for inferring stellar population parameters from photometry and spectroscopy spanning UV through IR wavelengths \citep{Johnson21}}}, the turnoff mass is around 1.7 $M_{\odot}$ \citep{Maeder09}. This corresponds well to a general remark by \cite{Longair98} that most of the light of  the old  stellar populations of galaxies comes from stars with a  mass in the range of $[1-2]$ M$_{\odot}$. 
A shift in mass by a factor 0.62  for JWST-ER1, as predicted by the scale-invariant theory,
implies a significant difference in the relative number frequency of stars as compared to the standard case.

Let us estimate the effect of such a shift on the average mass-luminosity ratio   $\langle M/L \rangle$ of a galaxy by some simple relations.
As an example, we take an initial mass function $\xi(\ln m)$  of the Salpeter form \citep{Salpeter55},
\begin{equation}
\xi(\ln m)= \frac{dN}{d\ln \, m}=\frac{dN}{dm} m= A m^{-x}, \quad \mathrm{or} \; \frac{dN}{dm} = A m^{-(1+x)},
\label{dnm}
\end{equation}
with $x=1.35$ (see Fig.~\ref{imf}). The mass of the galaxy then writes,
\begin{equation}
M_{gal}= \int \frac{dN}{dm} m \, dm= A \int m^{-x} dm=\;
\mid \frac{A}{-x+1} ({m_{m}}^{-x+1})\mid_{m_{min}}^{m_{max}}\;=A m_{m},
\label{mg}
\end{equation}
where $A$ is a normalization constant and where $m_{m}$ is the mean mass over the mass spectrum. This mean mass is $m_m= $13.7  M$_{\odot}$ for a mass interval that extends from $m_{min} = 0.01$  to $m_{max}=100$ M$_{\odot}$. If we call $\ell(m)$ the luminosity of a star of mass $m$,  the stellar  mass-luminosity relation writes, $\ell(m) =B \, m^{\alpha}$, where $B$ is another normalization constant.  The luminosity of a galaxy becomes,
\begin{equation}
L_{gal}= \int  \ell(m)\frac{dN}{dm} dm=A B \int m^{\alpha}\, m^{-(x+1)}dm = \;
\mid\frac{AB}{\alpha-x} {m}^{\alpha-x}\mid_{m_{min}}^{m_{max}}\;= AB \, m_L ,
\end{equation}
where $m_L$ is the mean mass over the luminosity distribution. Typical values are 1.7-1.8 M$_{\odot}$. Thus, the mean mass-luminosity ratio of a galaxy is scaling like,
\begin{equation}
\left \langle \frac{M}{L} \right \rangle = 
\frac{ (\alpha-x)}{B(1-x)} m_m^{1-x}  \; m_L^{x-\alpha} . 
\label{LM}
\end{equation}
If all the masses are reduced by the same factor, this factor applies both to $m_m$ and $m_L$. 
The values of $\alpha$ in the range of 1 to 2 M$_{\odot}$ considered earlier is quite large, hence making $\ell$ a very steep function of $m$, as shown in Fig.~\ref{imf}. We have, following \citet{Maeder09}
\begin{eqnarray} 
\ell  \sim  m^{\alpha},  & \hskip 5pt  \alpha=3.69,  & \hskip 5pt m \in [9 - 2]    \; M_{\odot}  \nonumber\\
                                        &  \hskip 5pt  \alpha=4.56,  & \hskip 5pt m \in [2 - 1]    \; M_{\odot}  \nonumber\\
                                        &  \hskip 5pt  \alpha=4.00,  & \hskip 5pt m \in [1 - 0.6] \; M_{\odot} 
\end{eqnarray}
%
%
The  individual stellar mass-luminosity ratios $m/\ell$  are varying  with  power $(1-\alpha)$ of the  mass $m$, while   Eq.~(\ref{LM}) shows that the factors are different when integrated over the mass spectrum of a galaxy. Now, we can see what is doing a change of the mass by a factor $f=0.62$ 
 (or 0.614 by \cite{Mercier24}), 
(this factor is the ratio $t/t_0= t$ corresponding to
  the reshift $z=1.94$ (or 2.02 by \cite{Mercier24})   of the lens in JWST-ER1). The ratio of the modified overall mass-luminosity ratio  to the standard unmodified case is,
\begin{equation}
\frac{\left \langle \frac{M}{L} \right \rangle_f}{\left \langle \frac{M}{L} \right \rangle_{\mathrm{std}} }  \sim  f^{1-x}  \; f^{x-\alpha} = f^{1-\alpha}.
\label{548}
\end{equation}
%
Interestingly enough, for a  change of all masses by the same factor $f$,  we come back to the same dependence  as 
given by the simple individual stellar expression  $(m/\ell) \sim m^{1-\alpha}$; 
for other changes of the mass distribution the full expression has evidently to be kept in.  
For of factor $f=0.62$  (or 0.614 by \cite{Mercier24})  and $\alpha=4.56$, 
 we get a value of  5.48  (or 5.67)  for the   ratio expressed in Eq. (\ref{548}). 
For  different average values of the  exponent of the M-L relation with $\alpha=4$,
 and 3\footnote{Such a value $\alpha=3$ could apply in case some star formation is still active in JWST-ER1.}, we get respectively  4.32  and 2.65 
with the redshift by Mercier et al.  
Thus,   whatever the average $\alpha$-value of the stellar population in JWSR-ER1, the real mass-luminosity ratio  may 
 be much higher  than the assumed one, which accounts for most  of the difference 
between the mass observed from lensing and the mass derived from the luminous matter. We recall that the  ratio of the total mass from lensing to
the stellar mass from photometry were  $\frac{M_{\mathrm{Tot}}}{M_*}$ was equal to about 5.9 by van Dokkum \cite{Dokkum23}
and about 2.7 by Mercier et al. \cite{Mercier24}. 


The so-called super-Salpeter IMF with $x \approx 1.65$ applied by \citet{Dokkum23} increases the total mass by a factor of about 4 compared to Salpeter's. In the scale invariant context, such a hypothesis of  a very bottom-heavy mass distribution, just opposed to the Chabrier one, is not necessary. 
 Despite the different remaining uncertainties in particular concerning the stellar population of the lens and the IMF, it appears that 
a standard mass distribution  can  naturally account  for the lensing mass without calling for dark matter. 
Thus, while the mass estimates by lensing are totally unaffected by scale invariant effects, it is not the case for the spectrophotometric determinations, 
which may be severely biased for distant galaxies, such as JWST-ER1.

\subsection{The SLACS sample of galaxies} 
\label{slacs}

To  complement  the above results for   JWST-ER1,  we turn towards the Sloan Lens ACS (SLACS) survey by Bolton \cite{Bolton2008} which is one of the largest and most homogeneous data set of galaxy-scale lenses, containing about 100 lenses analysed by Auger et al. \cite{Auger09}. One of the aim of the survey was to study  the distribution of luminous and dark matter out to redshift $z \approx 0.5$. It is based on multi-band imaging with ACS, WFC and NICMOS on the HST. The observations have been analysed with a lens model using an isothermal ellipsoid mass distribution, allowing high precision measurements of the mass within the Einstein radius for each lens.  The resulting mass estimates have been found unbiased compared to the estimates from SDSS photometry and are in the range $10^{10.5} < M < 10^{11.8}$ $M_{\odot}$, with a typical statistical error of 0.1 dex. The mean mass-luminosity ratio of the sample of the galaxies considered in  the SLACS sample is about $\langle M/L \rangle = 8.7$. Such a ratio  corresponds to a representative stellar mass of about 1.84 $M_{\odot}$, not much different  from the case of JWST-ER1. 

The survey provides for each lens the fractions $f^{Chab}_{*,Ein}$ and $f^{Salp}_{*,Ein}$ of the stellar mass within the Einstein radius for the Chabrier and Salpeter IMFs \cite{Chabrier03,Salpeter55} with respect to the total lensing mass. As stated by \cite{Auger09}, these fractions are independent of any priors from lensing, so that in some cases fractions higher than 1 were obtained. The survey provides a mean stellar mass fraction within the Einstein radius of $\langle f^{Chab}_{*,Ein}\rangle=0.4$ with a rms scatter of 0.1 for the Chabrier IMF; for the Salpeter's IMF the mean is $\langle f^{Salp}_{*,Ein}\rangle =0.7$ and a rms scatter of 0.2. 

In  the data of Table 3 and 4 by \cite{Auger09}, we select  three samples of lenses at different redshifts: 
for low redshifts, we select systems with $z_{\mathrm{L}} \leq 0.15$. For intermediate values, we take a sample
with redshifts bewteen 0.15 and 0.30, and for the highest range we consider galaxies 
  with $z_{\mathrm{L}}  \geq 0.30$, this last interval  although the largest one contains less galaxies than the other two.
This split in three bins   is a compromise between the need to have   different  mean redshifts
and also a sufficient number of galaxies in each sub-sample. The results are shown in Table~\ref{tab:SLACS}.  We see that the stellar mass fractions are 
  lower  for the higher  $z$-interval: compared to the lowest $z$-interval, the differences are  $\Delta f_* = 0.092$ for the Chabrier IMF
and  by $\Delta f_* = 0.166$ for the Salpeter IMF. Noticeably, the values of the stellar mass fractions for the intermediate zone
 lies inbetween the other two. 
These differences are rather small, of the same order as the uncertainties, unlike the high-$z$ case of JWST-ER1. 
The standard deviations of the different fractions are also given in Table~\ref{tab:SLACS}, allowing a test of significance.

Combining  the dispersions for the low and  the high $z$-intervals, 
we get a value $\sigma= 0.135$  for  the difference $ \Delta f_* = 0.449 -0.357= 0.092$  of the  $\langle f^{Chab}_{*,Ein}\rangle$ values. 
For a normal distribution, this corresponds to a probability of 50.4\% that the difference is significant.
  For $\langle f^{Salp}_{*,Ein}\rangle$, the difference is $\Delta f_* = 0.792-0.626=0.166$ and  $\sigma=0.239$, giving a probability of 51.2\%. 
Thus,  the differences are possible, but also well compatible with a random realization. 
The ages $t$ corresponding to the mean redshifts $z=0.114$, 0.218  and 0.383 of the samples  respectively 
are  $t=0.8591, 0.9113$ and 0.9502  in the dimensionless time scale, again for $\Omega_{\mathrm{m}}=0.05$. 
According to the modified conservation laws \cite{Maeder17a}, masses scale like $t$, 
which implies an increase  of the mass with decreasing redshifts.  The observations effectively suggest an increase 
of  the fractions $f^{Chab}_{*,Ein}$ and $f^{Salp}_{*,Ein}$. Let us make a comparison of observations and theory,   even
 if the observational uncertainties preclude from establishing this correspondence on a firm statistical ground.

As  in the case JWST-ER1, the Salpeter IMF gives masses in better agreement with observations than Chabrier's  IMF: the mass estimates from the photometric observations at $\langle z \rangle =0.114$ give a mass equal to 79.2\% of the total lensing mass. If we account, as above for JWST-ER1, for the shift of the Salpeter IMF resulting from the smaller masses in scale invariance, we need to rescale the stellar fraction in the Einstein radius following Eq.~(\ref{548}). For  the low redshift bin $\langle z \rangle=0.114$, we obtain today's corresponding value of the stellar mass fraction of $\langle f^{Salp}_{SIV}\rangle = 0.792 \times  0.9502^{-3.56}=  0.95$.  For the intermediate and high reshift bins, 
we obtain the fractions  $\langle f^{Salp}_{SIV}\rangle$ = 1.04 and 1.07. 
 These  fractions, very close to 1.0, mean that   when the shift in mass due to scale invariance is accounted for in absence of dark matter, 
the stellar masses from photometry/spectroscopy  within the Einstein ring, 
for a standard IMF, are remarkably close to the total masses from lensing, in support of the present interpretation.
\begin{table*}

\vspace*{0mm} 
 \caption{Analysis of the SLACS Survey: the mean stellar mass fractions within the Einstein radius, 
$\langle f^{Chab}_{*,Ein}\rangle$  and  $\langle f^{Salp}_{*,Ein}\rangle$  are given for  Chabrier and Salpeter IMFs 
respectively as well as the corresponding standard deviations among the two subsamples of redshifts. 
The column $\langle f^{Salp}_{SIV} \rangle $  gives the theoretical value  of the stellar mass fraction with Salpeter's IMF obtained after accounting
for the scale invariant correction. The last number gives the number of galaxies in each sub-sample. }
\label{tab:SLACS}
\begin{center} 
\scriptsize
\begin{tabular}{cccccccc}
\hline
$\mathrm{Redshift \; interval}$ &  $\langle z \rangle $  & $\langle f^{Chab}_{*,Ein}\rangle$  &  $\langle f^{Salp}_{*,Ein}\rangle$ &  
   $\sigma(f^{Chab}_{*,Ein})$   &$\sigma(f^{Salp}_{*,Ein})$ & $\langle f^{Salp}_{SIV} \rangle $ & $ N $\\
\hline\hline
$0  \leq z  \leq 0.15 $     &  0.114  & 0.449  & 0.792 & 0.110     & 0.197  & 0.95 & 17 \\
$0.15  \leq z  \leq 0.30 $ &  0.218  & 0.419  & 0.745 & 0.053     & 0.130  & 1.04 & 39 \\
$ 0.30 \leq z \leq 0.50   $&  0.383  & 0.357  &  0.626 & 0.079    & 0.136 & 1.07 & 10  \\
\hline
\end{tabular}
\end{center}
\normalsize
\end{table*}

\section{Discussion and proposed lensing tests}
\label{discussion}


Both the JWST-ER1 and the SLACS studies show that the measured differences between the stellar masses from photometry and from lensing are increasing with redshifts. We show that scale invariance theory reconciles the two mass estimates, both at low and high redshifts.  This is particularly stricking given the very large redshift difference between SLACS and JWST-ER1, and our finding applies as well within the SLACS sample that spans a limited redshift range.

Of note, different results obtained with GR and dark matter (DM) seem to severely contradict each other. At high redshift, like in JWST-ER1, the DM component vastly dominates the lens mass, to the point that a super-Salpeter IMF is invoked in addition to a DM halo with extreme densities. Recently \citet{Kong24} attempted to alleviate the DM density by invoking the presence of self-interacting DM. But their overall conclusion that JWST-ER1 is an extreme lens in terms of DM content at high redshift comes in contrast with several recent studies on rotation curves of galaxies by Nestor Shachar \cite{Nestor Shachar23} that corroborate previous results by the Genzel group, showing that much less dark matter is needed in GR at high reshift than at low redshift. For example, only a fraction of about 17\% of the typical value appears to be remaining in galaxies at $z=2.44$ \cite{Nestor Shachar23}, in sharp contrast with JWST-ER1. 



The present work focuses on the comparison of the total lensing mass of galaxies and the total luminous mass in GR and in scale invariance. It suggests that, in scale invariance, the total luminous mass is sufficient to explain the observed deflection angle without the need of dark matter, given the observational errors. This paves the way for several other more detailed tests involving lensing in scale invariance:

\begin{itemize}

\item {\bf Pixel-level modeling:} state-of-art modeling of lens galaxies decomposes their mass into luminous mass plus dark matter NFW halo \cite[e.g.][]{Suyu2013, Suyu2014}. This allows to model HST images of lensing systems down to the noise level. A first test for scale invariance lensing would be to attempt the same without the DM component, just using the light map of the lens and converting it into mass following the recipes of the present work.
 Lensing constrains the lensing potential only locally, in the close vicinity of the multiple images as emphasized by Wagner \cite[]{Wagner18, Wagner19, Wagner22}. In this context,  it will be interesting to see if lens potentials with little or even no dark matter in SIV can predict the same local lens properties than standard GR models with dark matter. We note that such a test will require to map as much as possible of the baryonic mass of the lens, including the gas and dust, with deep infrared and radio observations with JWST and ALMA. Missing direct gas and dust components may mimic the same degeneracies as missing dark matter, although at a much lower extend. 

\item {\bf Lens dynamics:} dynamical information is available for some strong lens systems, including SLACS. Joint lens/dynamical analysis of strong lenses in the context of scale invariance would further test the theory. This is not straightforward, however, as the virial theorem does not apply directly in scale invariance and requires to account for the additional acceleration due to scale invariance. In an analytical study of the dynamics of clusters of galaxies in the scale invariant context \cite{Maeder24a}, we  searched for an integral of the equation of motion using adiabatic invariants. The result was a different relation between the velocity dispersion and the mass of the system a compared with standard GR. In SIV a smaller mass is associated to the lens for a given velocity dispersion, supporting further the present result from gravitational lensing and calling for a joint lensing/dynamical analysis. Finally, the dynamical analysis, especially if conducted with spatially-resolved data, will constrain the lens mass profile at positions not neces\-sarily constrained by lensing alone \cite{Yildirim23, Shajib23}. Such observations can be obtained with JWST and NIRSpec IFU or, soon, with the Extremely Large Telescope and adaptive optics.

\item {\bf Time delay cosmography:} Time delays in lensed quasars can be converted into an $H_0$ measurement,  provided that the local lensing potential at the multiple image is well described as emphasized by \cite{Wagner18} and by  \cite{SS13, SS14}. This local potential must also be compatible with the rest of the lens galaxy.  In the scale invariance context, the absence of dark matter removes a well known dominant source of degeneracy \cite{Wagner18}  between the source and lens properties, which is an advantage for $H_0$ determinations.
Evidently, some source of degeneracy may be remaining, such as the gas. While challenging to observe, it is still easier than measuring dark matter. It will be interesting to see which values of $H_0$ are favored by scale-invariance lens models.  Current lens models with dark matter tend to favor $H_0$ values compatible with the local measurements with Cepheids \cite[e.g.][]{Riess22}. 


\item {\bf Lensing by edge-on spirals:} The luminous mass in edge-on galaxies is of course extremely elliptical. In scale invariance the mass is dominated by baryons (or even fully accounted for by baryons) and therefore the difference between the configurations of lensed images produced no-DM scale invariance models and standard models dominated by a fairly spherical halo, should offer great discriminating power between the two theories. In fact, hints of such behaviour may have been found already, for example with the spectacular case of J220132.8-320144 where some of the lensed images predicted by standard models with DM are missing from HST imaging data \cite{Chen2013}. 
The gas and dust components in edge-on spirals are even more important than in other more massive and passive galaxies like SLACS. However, those can be mapped in the IR and radio and gas and dust are mostly in the disk of the galaxy rather than the halo. Gas and dust therefore do not alter the above statement that lensed images produced by a spherical DM halo will differ significantly from lensing by just baryons in the disk and bulge, whether they be stars, gas or dust. We consider lensing by edge-on spirals as one of the most powerful discriminating tests between SIV and stardard GR. 

\item {\bf Weak lensing:} Weak lensing mass maps of galaxy clusters are unaffected by scale invariance, just as for strong lensing. However, as for strong lensing, scale invariance changes the infered luminous mass. It will therefore be crucial to compare the weak lensing mass infered for galaxy clusters to that infered from spectroscopy for all galaxy members, for example with ultra deep integral field spectroscopy. Additionally, and as shown in \citep{Maeder24a}, the scale invariance mass infered from galaxy velocities also has to be consistent with the scale-invariant luminous mass.

\end{itemize}

The above proposed tests are beyond the scope of the present paper but most observational material to implement them do exist and will be used in the near future. 

\section{Conclusions}  
\label{concl}
Detailed calculations  of the geodesic equation to study deviation of light  rays passing a massive object have been developed  in the scale invariant theory. The presence  of a lensing galaxy and scale invariant effects  generate two additional terms in the equation of the geodesics with respect to General Relativity. However, in practice, numerical integration of the geodesics performed specifically for the  Einstein ring JWST-ER1 with its lensing galaxy at $z_L \cong 2$ indicates that the change in deflection angle is typically a fraction $\mathcal{O}$($10^{-6}$) of that predicted by GR. The lensing deflection can therefore be considered as independent from scale invariant effects. However, while scale invariance has no effect on the photons' geodesics themselves, its actual impact resides in the interpretation and use of the mass-luminosity relation of the stellar population in the lens galaxy and of its IMF. In this context, the need to call for dark matter does not appear necessary.

\acknowledgments
Both authors are grateful to the referee for correctly pointing out some critical points to be solved in the demonstation. A.M.  expresses his best  thanks to Dr. Vesselin Gueorguiev for constructive interaction since many  years, and is most grateful to Prof. James Lequeux for shis support and encouragements.






\end{document}